\documentclass[paper]{jhep3} 

\JHEPspecialurl{http://jhep.sissa.it/JOURNAL/JHEP3.tar.gz}

\usepackage{epsfig,multicol}
\usepackage{cite}

\usepackage{amsmath}
\usepackage{amssymb}
\preprint{
KEK-TH-968 \\
KUNS-1925 \\
RIKEN-TH-39 \\
IU-TH-2 \\
hep-th/0506205\\}

\title{Perturbative versus nonperturbative dynamics \\
of the fuzzy S$^{2} \times$ S$^{2}$}
\author{ Takehiro Azuma${}^a$, Subrata Bal${}^{b,c}$, 
Keiichi Nagao${}^{a,d}$ and Jun Nishimura${}^{a,e}$ \\
\llap{$^a$}High Energy Accelerator Research Organization (KEK),\\
1-1 Oho, Tsukuba 305-0801, Japan  \\
\llap{$^b$}Department of Physics, Kyoto University,\\
Kitashirakawa, Kyoto 606-8502, Japan\\
\llap{$^c$}Theoretical Physics Laboratory,\\
The Institute of Physical and Chemical Research (RIKEN), \\
2-1 Hirosawa, Wako, Saitama 351-0198, Japan\\
\llap{$^d$}Theoretical Physics Laboratory, College of Education,
Ibaraki University, \\ 
2-1-1 Bunkyo, Mito, Ibaraki 310-8512, Japan \\
\llap{$^e$}Department of Particle and Nuclear Physics,\\
Graduate University for Advanced Studies (SOKENDAI),\\
1-1 Oho, Tsukuba 305-0801, Japan \\
\email{azumat@post.kek.jp, 
subrata@riken.jp, 
nagao@mx.ibaraki.ac.jp, 
jnishi@post.kek.jp}} 

\abstract{We study a matrix model with a cubic term,
which incorporates
both the fuzzy $\stwostwo$ and the fuzzy $\stwo$ as classical solutions.
Both of the solutions
decay into the vacuum of the pure Yang-Mills model
(even in the large-$N$ limit) when the coefficient
of the cubic term is smaller than a critical value, but the 
large-$N$ behavior of the critical point is different for
the two solutions. 
The results above the critical point are nicely reproduced by
the all order calculations in perturbation theory.
By comparing the free energy,
we find that the true vacuum is given either by 
the fuzzy $\stwo$ or by the ``pure Yang-Mills vacuum'' depending on the
coupling constant.
In Monte Carlo simulation we do observe a decay of the
fuzzy $\stwostwo$ into the fuzzy $\stwo$ at moderate $N$,
but the decay probability seems to be suppressed at large $N$.
The above results, together with our previous results 
for the fuzzy CP$^2$, reveal certain universality in
the large-$N$ dynamics of four-dimensional fuzzy manifolds 
realized in a matrix model with a cubic term.
}


\newcommand{\bel}{\begin{equation}\label}

\newcommand{\non}{\nonumber \\}
\newcommand{\n}{\nonumber}
\newcommand {\beq}{\begin{equation}}
\newcommand {\eeq}{\end{equation}}
\newcommand {\beqa}{\begin{eqnarray}}
\newcommand {\eeqa}{\end{eqnarray}}
\newcommand {\bc}{\begin{center}}
\newcommand {\ec}{\end{center}}
\newcommand {\tr}{{\rm tr\,}}

\newcommand {\ee}{\mbox{e}}

\newcommand {\dd}{\mbox{d}}

\newcommand {\defeq}{\stackrel{\rm def}{=}}

\newcommand {\stwo}{{\rm S}^{2}}
\newcommand {\stwostwo}{ {\rm S}^{2} \times {\rm S}^{2} }

\def\dag{\dagger}

\def\vs5{\vspace*{5mm}}
\def\vs1{\vspace*{1cm}}
\def\vs2{\vspace*{2cm}}
\def\hs5{\vspace*{5mm}}
\def\hs1{\hspace*{1cm}}
\def\hs2{\hspace*{2cm}}
\def\vs50{\vspace*{50mm}}
\def\vs20{\vspace*{20mm}}

\def\tr{\hbox{tr}}

\begin{document}

\section{Introduction and summary}
Fuzzy spheres \cite{Madore}, 
or fuzzy manifolds in general,
have recently
attracted much attention in various branches of particle physics.
One of the motivations comes from the general expectation 
that the noncommutative geometry, which is characteristic
to those manifolds, provides
crucial links to string theory and quantum gravity.
Indeed Yang-Mills theories on noncommutative geometry
appear in a certain low energy limit of string theory \cite{Seiberg:1999vs}.
There is also an independent observation that the space-time 
uncertainty relation, which is naturally realized by noncommutative
geometry, can be derived from some general assumptions
on the underlying theory of quantum gravity \cite{gravity}.
Another motivation is to use fuzzy manifolds
as a regularization scheme alternative to 
the lattice regularization \cite{Grosse:1995ar}.
Unlike the lattice, fuzzy manifolds typically
preserve the continuous symmetries 
of the space-time considered, and hence it is expected that 
the situation concerning chiral symmetry 
\cite{Grosse:1994ed,%
Carow-Watamura:1996wg, chiral_anomaly, non_chi,balagovi,%
chiral_anomaly2,balaGW,Nishimura:2001dq,AIN,AIN2,Ydri:2002nt,%
Iso:2002jc,Balachandran:2003ay,nagaolat03, AIN3} 
and supersymmetry might be improved.

As expected from the connection to string theory \cite{Myers:1999ps},
fuzzy manifolds appear as classical solutions
in matrix models with a Chern-Simons-like term
\cite{0003187,0101102,0204256,0207115,0301055}
and their dynamical properties have been studied
in refs.\ \cite{0108002,0206075,%
0303120,0307075,0309082,%
0312241,0402044,0403242,0407089,0412052,0412312,0503041,0506033,0506044}. 
One can actually use matrix models to define a regularized 
field theory on the fuzzy spheres as well as on a noncommutative 
torus \cite{AMNS}, which enables nonperturbative studies of such
theories from first principles \cite{simNC}. 
These matrix models belong to the class of the 
so-called dimensionally reduced models
(or large-$N$ reduced models),
which is widely believed to provide a constructive definition of 
superstring and M theories \cite{9610043,9612115}.
In fact there are certain evidences in the IIB matrix model \cite{9612115}
that {\em four-dimensional} space-time is generated dynamically 
  \cite{Nishimura:2001sx,KKKMS,Kawai:2002ub,0307007}.
In refs.~\cite{Nishimura:2001sx,KKKMS,Kawai:2002ub} 
the free energy of space-time
with various dimensionality has been calculated
using the gaussian expansion method, and the free energy turned out to
take the minimum value for the four-dimensional space-time.
In ref.~\cite{0307007} it was found that
the fuzzy ${\rm S}^2 \times {\rm S}^2$ 
(but not the fuzzy ${\rm S}^2$) is a solution
to the 2-loop effective action.
See also refs.\ \cite{Aoki:1998vn,9811220,Ambjorn:2000bf,%
Ambjorn:2000dx,NV,Burda:2000mn,%
Ambjorn:2001xs,exact,sign,Vernizzi:2002mu,%
Nishimura:2004ts,Bal:2004ai,Nishimura:2003rj}
for related works on this issue.
The fuzzy sphere is also useful \cite{Aschieri:2003vy}
in the Coset Space Dimensional Reduction
\cite{Forgacs:1979zs,Kapetanakis:1992hf}.


In view of the variety of contexts in which fuzzy manifolds
have been studied,
we consider it important to study 
their nonperturbative dynamics from first principles
by Monte Carlo simulations.
In ref.\ \cite{0401038} we have studied
the dimensionally reduced 3d Yang-Mills-Chern-Simons model, 
which incorporates the fuzzy S$^{2}$
as a classical solution \cite{0101102}.
We have observed a first-order phase transition as we vary 
the coefficient ($\alpha$) of the Chern-Simons term.
For small $\alpha$
the large-$N$ behavior of the model is the same as in the 
pure Yang-Mills model, whereas for large $\alpha$
a single fuzzy S$^2$ appears dynamically.
The emergence of a fuzzy sphere in matrix models
may be regarded as a prototype of the dynamical generation of
space-time since it has lower dimensionality
than the original dimensionality that the model can actually describe.
In ref.\ \cite{0410263} we have performed the all order calculations
in perturbation theory around the fuzzy S$^2$ solution
following the proposal in ref.\ \cite{0403242},
and confirmed that the Monte Carlo results for various observables
in the ``fuzzy sphere phase''
can be nicely reproduced.
For obvious reasons it is interesting to extend these works to
{\em four}-dimensional fuzzy manifolds.
In ref.\ \cite{0405096} we have studied a matrix model
incorporating the fuzzy S$^{4}$, and
in ref.\ \cite{0405277} a matrix model, 
which incorporates the fuzzy CP$^2$ as well as the fuzzy S$^2$.
%

In this paper we study the fuzzy $\stwostwo$,
which has been studied extensively in the literature 
\cite{0307007,0312241,0403242,0412312,0503041,0506033,0506044}.
We perform perturbative and nonperturbative studies
of a 6d Yang-Mills model with a cubic term,
which incorporates the fuzzy $\stwostwo$ as well as
the fuzzy $\stwo$ as classical solutions.
Both of the solutions become
unstable (even in the large-$N$ limit) when the coefficient of
the cubic term is smaller than a critical value, but the 
large-$N$ behavior of the critical point is different for
the two solutions. 
The results above the critical point are nicely reproduced by
the all order calculations in perturbation theory.
By comparing the free energy,
we find that the true vacuum is given either by 
the fuzzy $\stwo$ or by the pure Yang-Mills vacuum depending on the
coupling constant.
In Monte Carlo simulation we do observe a decay of the
fuzzy $\stwostwo$ into the fuzzy $\stwo$ at moderate $N$,
but the decay probability seems to be suppressed at large $N$.

In fact the above results are qualitatively similar
to the fuzzy CP$^2$ case
as far as large-$N$ properties are concerned.
This reveals certain universality in
four-dimensional fuzzy manifolds realized in a matrix 
model with a {\em cubic} term.
In the case of the fuzzy S$^{4}$ \cite{0405096}
we had to add a {\em quintic} term, which led to a totally
different situation.
We note, however, that even within matrix models with a cubic term,
we have found 
two directions which lead to qualitatively different results.
In ref.\ \cite{0504217} it is shown that 
an additional ``mass term'' can make various fuzzy sphere solutions
almost degenerate at the classical level, and it is possible that 
the true quantum vacuum is described by a set of coincident fuzzy spheres,
giving rise to nontrivial gauge groups.
In ref.\ \cite{0506062} we have shown that supersymmetry removes
the quantum effects, and as a result the single fuzzy sphere becomes
always stable if the large-$N$ limit is taken in such a way that
various correlation functions scale.

This paper is organized as follows.
In section \ref{model} 
we define the model and discuss its classical solutions.
In sections \ref{stwostwo} and \ref{stwo} we study the properties of the
fuzzy $\stwostwo$ and the fuzzy $\stwo$, respectively,
by performing Monte Carlo simulations and the 
all order calculations in perturbation theory.
In section \ref{true} we determine the true vacuum of the model
based on the comparison of the free energy.
In section \ref{decay} we discuss a transition from the fuzzy
$\stwostwo$ to the fuzzy $\stwo$ observed in Monte Carlo simulation.
In the appendices we provide the details of our calculations.

Note added: 
Part of this work has been reported by one of the authors
at YITP workshop ``Quantum Field Theory 2004'', July, 2004,
at a meeting of the Physical Society of Japan,
Kochi, Sep.\ 2004 and
at Dublin Institute for Advanced Studies, Dec.\ 2004.
While we were preparing this article, 
we received a preprint \cite{0506044}, which discusses
the phase diagram of a similar model 
based on the one-loop effective action.



 
\section{The model and its classical solutions}
\label{model}
  The model we study in this paper is defined by the action
  \begin{eqnarray}
   S[A] = N \, \textrm{tr} \left( - \frac{1}{4} \, 
   [A_{\mu}, A_{\nu}]^2 + \frac{2}{3} \, i \, 
  \alpha \, f_{\mu \nu \rho} \,  A_{\mu} A_{\nu} A_{\rho} 
  \right) \ , \label{s2s2action} 
  \end{eqnarray}
  where $A_{\mu}$ ($\mu = 1, \cdots , 6$) 
  are traceless $N \times N$ hermitian matrices.
  Here and henceforth we assume that
 repeated Greek indices are summed over all possible integers.
  The rank-3 tensor $f_{\mu \nu \rho}$ in the cubic term is given by
   \begin{eqnarray}
     f_{\mu \nu \rho} = \left\{ \begin{array}{ll}
     \epsilon_{\mu \nu \rho} & 
     \mbox{~~~for~$\mu, \nu, \rho = 1,2,3$} \ , \\
     \epsilon_{\mu \nu \rho} & 
     \mbox{~~~for~$\mu, \nu, \rho = 4,5,6$}  \ , \\
     0 & \mbox{~~~otherwise} \ ,
   \end{array} \right.  \label{s2s2structure}
  \end{eqnarray}
  where $\epsilon_{\mu \nu \rho}$ is a totally antisymmetric tensor
  with $\epsilon_{123} = \epsilon_{456} = 1$.

  The classical equation of motion reads
  \begin{eqnarray}
   [A_{\nu}, [A_{\mu}, A_{\nu}]] - i \, \alpha \, f_{\mu \nu \rho} \, 
   [A_{\nu}, A_{\rho}] = 0 \ .
  \end{eqnarray} 
  In addition to the commutative solution, which exists also 
  for $\alpha = 0$, it has the fuzzy $\stwostwo$ solution, 
  which is defined by
   \begin{eqnarray}
   A_{\mu} = X_\mu^{(n_1,n_2;k)}
 \defeq
 \left\{ \begin{array}{ll} 
     \alpha \, \Bigl(L^{(n_{1})}_{\mu} \otimes {\bf 1}_{n_{2}}\Bigr) \otimes 
     {\bf 1}_{k} &  \mbox{~~~for~$\mu = 1,2,3$} \ , \\ 
     \alpha \, \Bigl({\bf 1}_{n_{1}} \otimes L^{(n_{2})}_{\mu} \Bigr) \otimes
     {\bf 1}_{k} &  \mbox{~~~for~$\mu = 4,5,6$} \ ,
   \end{array} \right. \label{s2s2fuzzysphere}
  \end{eqnarray}
where $L^{(n)}_{\mu}$ represents the $n$-dimensional representation
of the SU$(2)$ Lie algebra, and the integers $n_1$, $n_2$ and $k$
should satisfy $N = n_{1} \, n_{2} \, k$.
Let us define the ``Casimir operators''
  \begin{eqnarray}
   Q_{1} = \sum_{\mu=1}^{3} (A_{\mu})^{2} \ , \hspace{2mm}
   Q_{2} = \sum_{\mu=4}^{6} (A_{\mu})^{2} \ .
  \label{s2s2casimirdef}
  \end{eqnarray}
 Plugging the solution (\ref{s2s2fuzzysphere}), we obtain
 $Q_j = R_j {\bf 1}_N $, where
 $R_j= \frac{1}{2} \alpha \sqrt{ (n_j) ^2 -1}$
 for $j=1,2$.
 This classical solution (\ref{s2s2fuzzysphere}) therefore
  describes $k$ coincident fuzzy $\stwostwo$
 with the radii $R_1$ and $R_2$ in the $123$- and $456$- directions, 
 respectively.
 If we expand the model around the solution, we obtain
 the U($k$) gauge theory on the fuzzy $\stwostwo$ generalizing
 the work on the fuzzy $\stwo$ \cite{0101102}.
 The classical action for the solution is given by
\begin{eqnarray}
 S[X^{(n_1,n_2;k)}]
  = - \frac{N^{2} \alpha^{4}}{24} 
 \Bigl\{ (n_{1})^{2} + (n_{2})^{2} - 2 \Bigr\} \ .
\label{cl-action}
\end{eqnarray}
 In what follows we focus on the symmetric fuzzy $\stwostwo$ 
 ($n_{1} = n_{2}$) and the fuzzy $\stwo$ ($n_{2} = 1$).
 Monte Carlo simulation is performed
 using the heat bath algorithm as in refs.\ \cite{9811220,0401038}.


\FIGURE{
    \epsfig{file=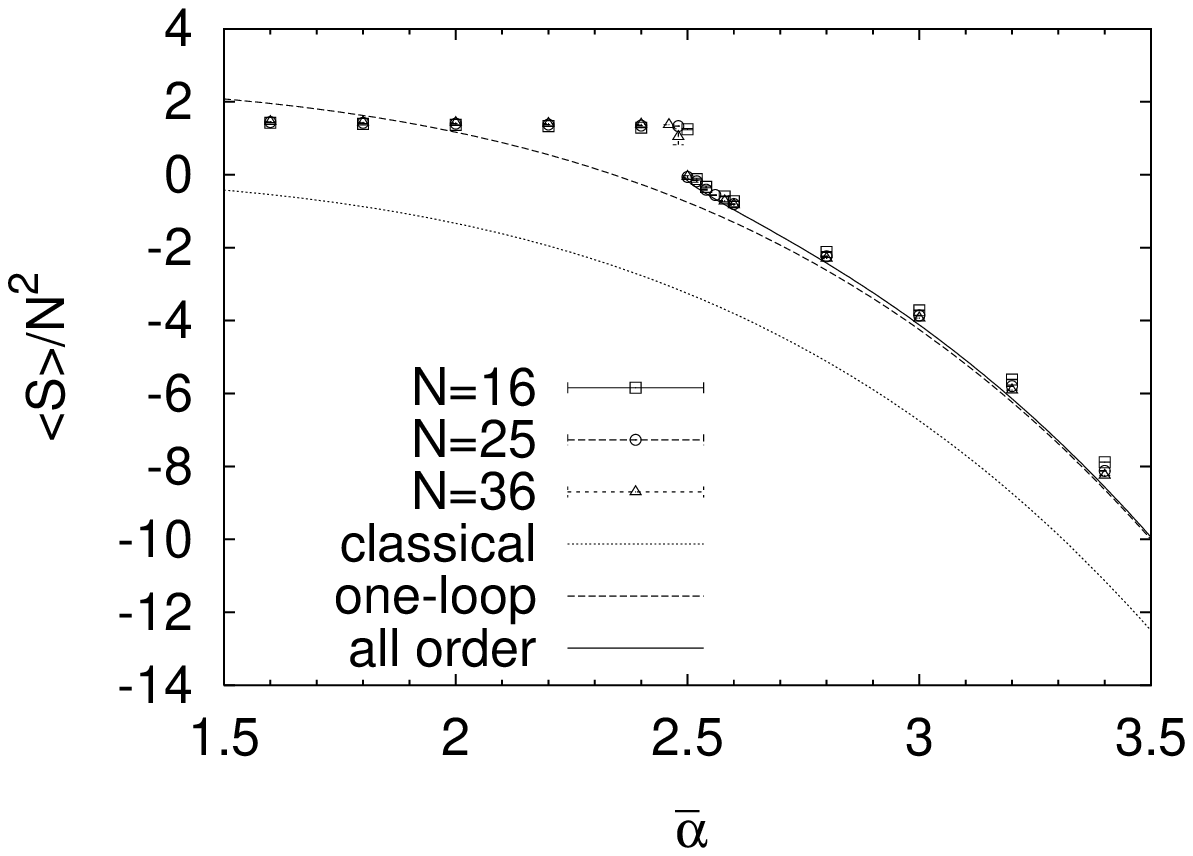,width=7.4cm}
    \epsfig{file=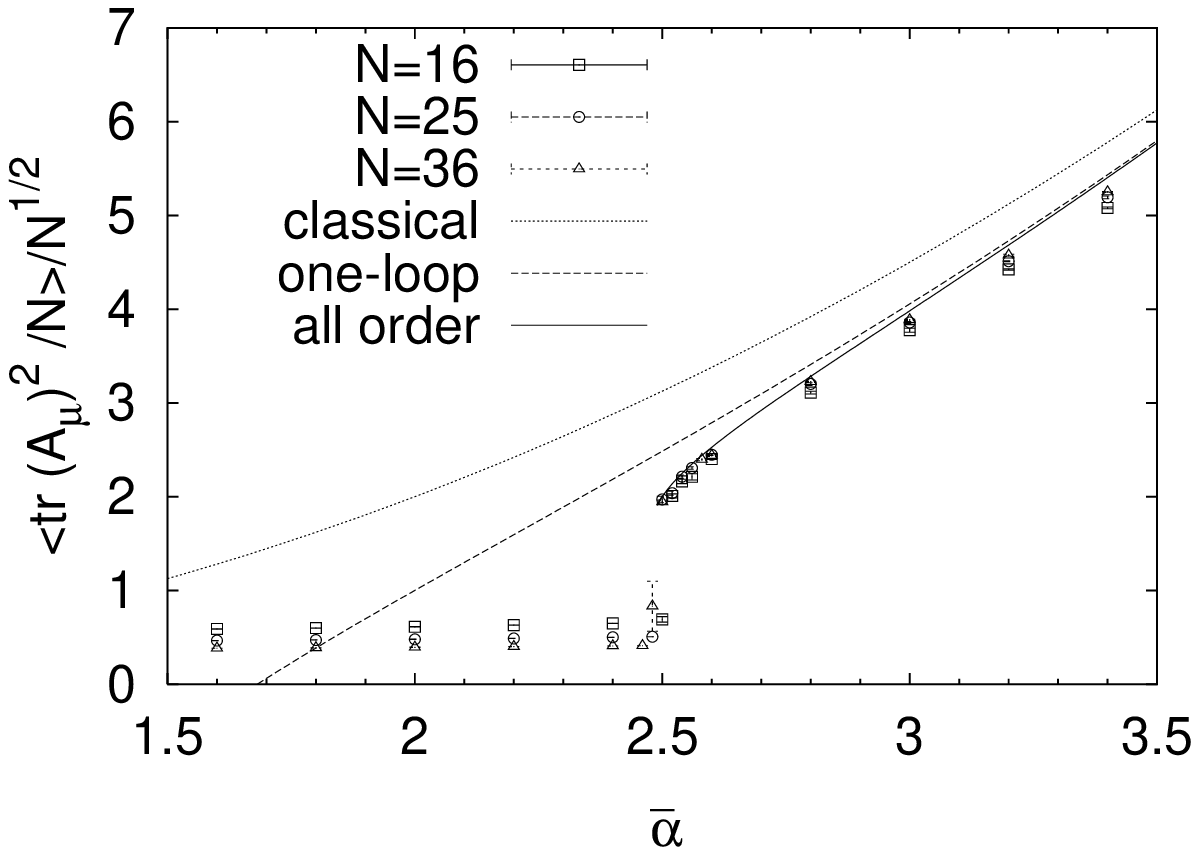,width=7.4cm}
   \caption{The observables obtained by Monte Carlo simulation
 with the fuzzy $\stwostwo$ start
 are plotted against ${\bar \alpha} =
 \alpha N^{\frac{1}{4}}$ for $N=16, 25, 36$ (i.e., $n=4,5,6$). 
 The dotted (dashed) lines
 represent the classical (one-loop) results at large $N$. 
 The solid lines represent the all order results at large $N$
 obtained above the critical point.} 
 \label{miscS2S2FS} 
}

\section{Properties of the fuzzy $\stwostwo$}
\label{stwostwo}

In this section we study the properties of the single ($k=1$)
fuzzy $\stwostwo$.
We perform Monte Carlo simulation 
taking $A_{\mu} = X_\mu^{(n,n;1)}$ as the initial configuration,
where $n=\sqrt{N}$.
Let us consider the expectation value of the action (\ref{s2s2action})
and the ``extent of space-time''
$\frac{1}{N} \, \tr (A_\mu)^2 = \frac{1}{N} \, \tr \, (Q_1+Q_2)$.
In figure \ref{miscS2S2FS} we plot these quantities
obtained by the simulation with $N=16,25,36$ against 
$\bar \alpha =  \alpha \, N^{\frac{1}{4}}$.
We observe a discontinuity at ${\bar \alpha} \simeq 2.5$, which implies
the existence of a first-order phase transition.
The critical point agrees with the result
(\ref{kuri}) obtained analytically from the 
effective action.

Above the critical point we can calculate 
the observables
in the large-$N$ limit
to all orders in perturbation theory as
 \begin{eqnarray}
   \frac{1}{N^{2}} \langle S \rangle &\simeq& 
   - \frac{{\bar \alpha}^{4}}{12} + \frac{5}{2}
  + \frac{8}{{\bar \alpha}^{4}} + \frac{448}{3{\bar \alpha}^{8}}
  + \frac{3520}{{\bar \alpha}^{12}} + \cdots \ ,
    \label{s2s2actok1loop} \\
   \frac{1}{\sqrt{N}} \left\langle \frac{1}{N} \tr (A_\mu)^2 \right\rangle
   &\simeq&  \frac{{\bar \alpha}^{2}}{2} - \frac{4}{{\bar \alpha}^{2}}
   - \frac{40}{{\bar \alpha}^{6}}
   - \frac{768}{{\bar \alpha}^{10}} - \frac{18304}{{\bar \alpha}^{14}}
   - \cdots \ .
  \label{s2s2a-sqk1loop} 
\end{eqnarray}
(See appendix \ref{allorder-s2s2} for derivation.)
This result, as well as the classical and one-loop results,
is plotted in figure \ref{miscS2S2FS}.
We observe that the Monte Carlo data for 
$\bar \alpha > \bar {\alpha}_{\rm cr}$ approach
the all order results as $N$ increases.

  \FIGURE{
    \epsfig{file=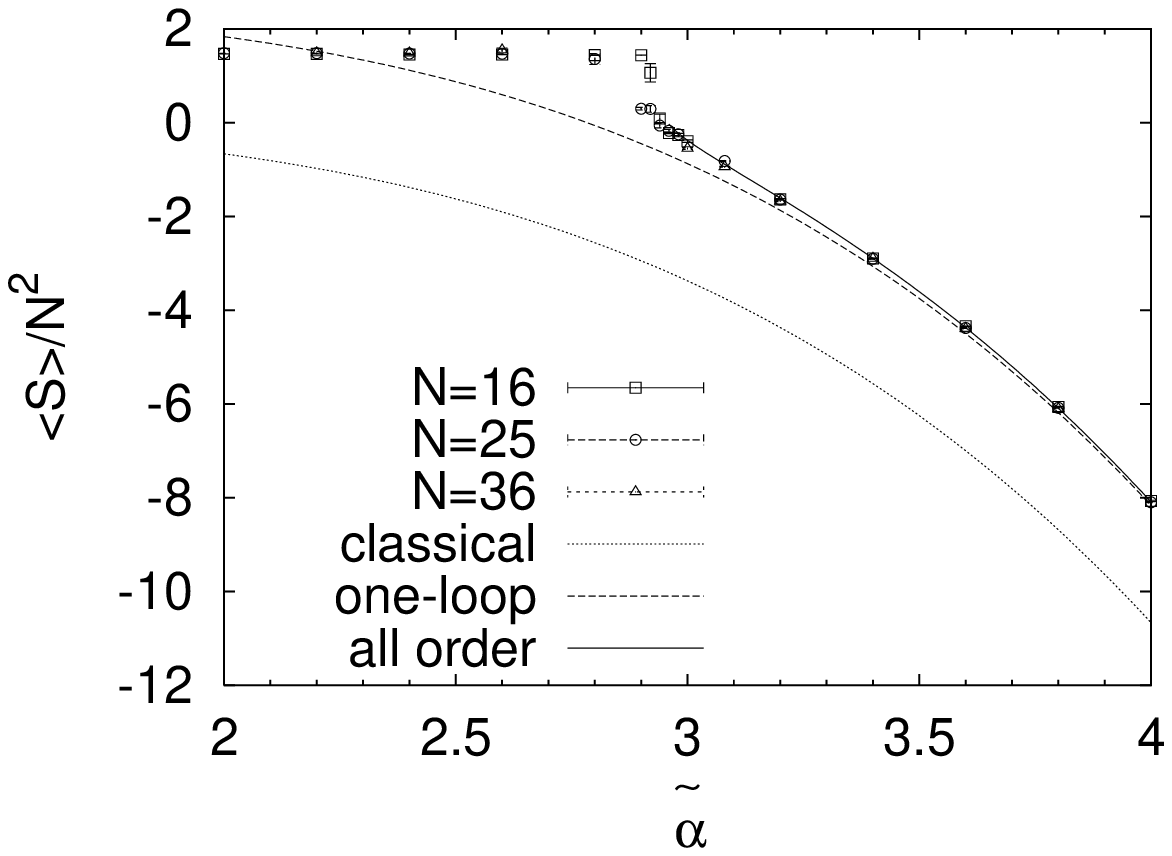,width=7.4cm}
    \epsfig{file=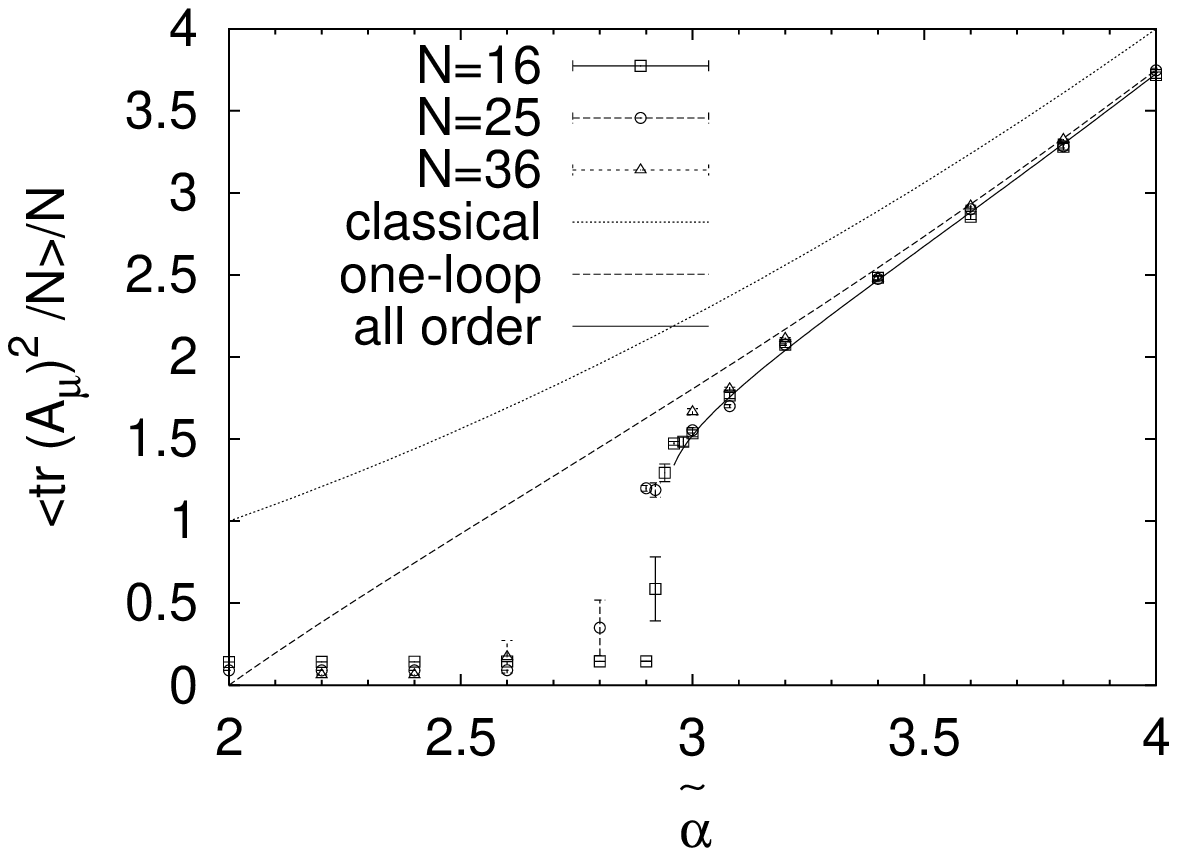,width=7.4cm}
   \caption{
The observables obtained by Monte Carlo simulation
 with the fuzzy $\stwo$ start
 are plotted against ${\tilde \alpha} =
 \alpha N^{\frac{1}{2}}$ for $N=16, 25, 36$.
The dotted (dashed) lines
 represent the classical (one-loop) results at large $N$.
 The solid lines represent the all order results at large $N$
 obtained above the critical point.} 
 \label{miscS2S2-S2FS} }

\section{Properties of the fuzzy $\stwo$}
\label{stwo}

Next we study the properties of the single ($k=1$) fuzzy $\stwo$.
We perform Monte Carlo simulation
taking $A_{\mu} = X_\mu^{(N,1;1)}$ as the initial configuration.
In figure \ref{miscS2S2-S2FS} we plot the results
for $N=16, 25, 36$ 
against ${\tilde \alpha} = \alpha N^{\frac{1}{2}}$.
We observe a gap at the
critical point ${\tilde \alpha}  \simeq 3.0$, 
which agrees with the result (\ref{anal-crit_S2})
obtained from the effective action.
The Monte Carlo results above this point
agree well with the all order results in perturbation theory 
at large $N$, which are given by
  \begin{eqnarray}
   \frac{1}{N^{2}} \langle S \rangle &\simeq&
    - \frac{{\tilde \alpha}^{4}}{24} + \frac{5}{2}
  + \frac{16}{{\tilde \alpha}^{4}} + \frac{1792}{3{\tilde \alpha}^{8}}
  + \frac{28160}{{\tilde \alpha}^{12}} + \cdots \ , 
   \label{s2s2-s2acto1loop} \\
   \frac{1}{N} \left\langle \frac{1}{N} \tr (A_\mu)^2 \right\rangle
   &\simeq&  \frac{{\tilde \alpha}^{2}}{4} - \frac{4}{{\tilde \alpha}^{2}}
  - \frac{80}{{\tilde \alpha}^{6}} - \frac{3072}{{\tilde \alpha}^{10}}
  - \frac{146432}{{\bar \alpha}^{14}} - \cdots \ .
  \label{s2s2-s2a-sq1loop}
  \end{eqnarray}
(See appendix \ref{allorder-s2s2-s2} for derivation.)




\section{The true vacuum}
\label{true}

In this section we determine
the ``true vacuum'' by comparing the free energy,
which can be obtained to all orders in perturbation theory as
  \begin{eqnarray}
\frac{1}{N^2} \,   W_{k \, \stwostwo} &\simeq& 
- \frac{{\bar \alpha}^{4}}{12k} 
+ 4 \log {\bar \alpha}
 + \log \frac{16N^4}{k^2} - \frac{15}{4} 
 - \frac{8k}{{\bar \alpha}^{4}} 
  - \frac{224k^{2}}{3{\bar \alpha}^{8}}
  - \cdots
 \ , \label{s2s2eff1loop} \\
\frac{1}{N^2} \,    W_{k \, \stwo} &\simeq&  
  - \frac{{\tilde \alpha}^{4}}{24k^{2}} 
  + 4 \log {\tilde \alpha}
  +  \log{\frac{N^5}{k^4}} - \frac{11}{4}
  - \frac{16k^{2}}{{\tilde \alpha}^{4}} 
  - \frac{896k^{4}}{3{\tilde \alpha}^{8}}
  - \cdots    \label{s2s2-s2eff1loop} 
  \end{eqnarray}
for the $k$ coincident fuzzy $\stwostwo$
and the $k$ coincident fuzzy $\stwo$, respectively.
 (See appendices \ref{allorder-s2s2} and
 \ref{allorder-s2s2-s2} for derivation.)
Note that these solutions are stable
above the critical points given by
(\ref{kuri}) and (\ref{anal-crit_S2}).

 Let us first compare the free energy among different values of $k$.
 In figure \ref{weff-k} we plot the free energy
 for the fuzzy $\stwostwo$ (left) and $\stwo$ (right)
 against $\bar \alpha$ and $\tilde \alpha$, respectively, for various $k$.
 The free energy for $k > 1$ is always larger
 than that for $k=1$, which implies that
 the dynamically generated gauge group is U$(1)$
 for both the fuzzy $\stwostwo$ and the fuzzy $\stwo$ cases.

   \FIGURE{
    \epsfig{file=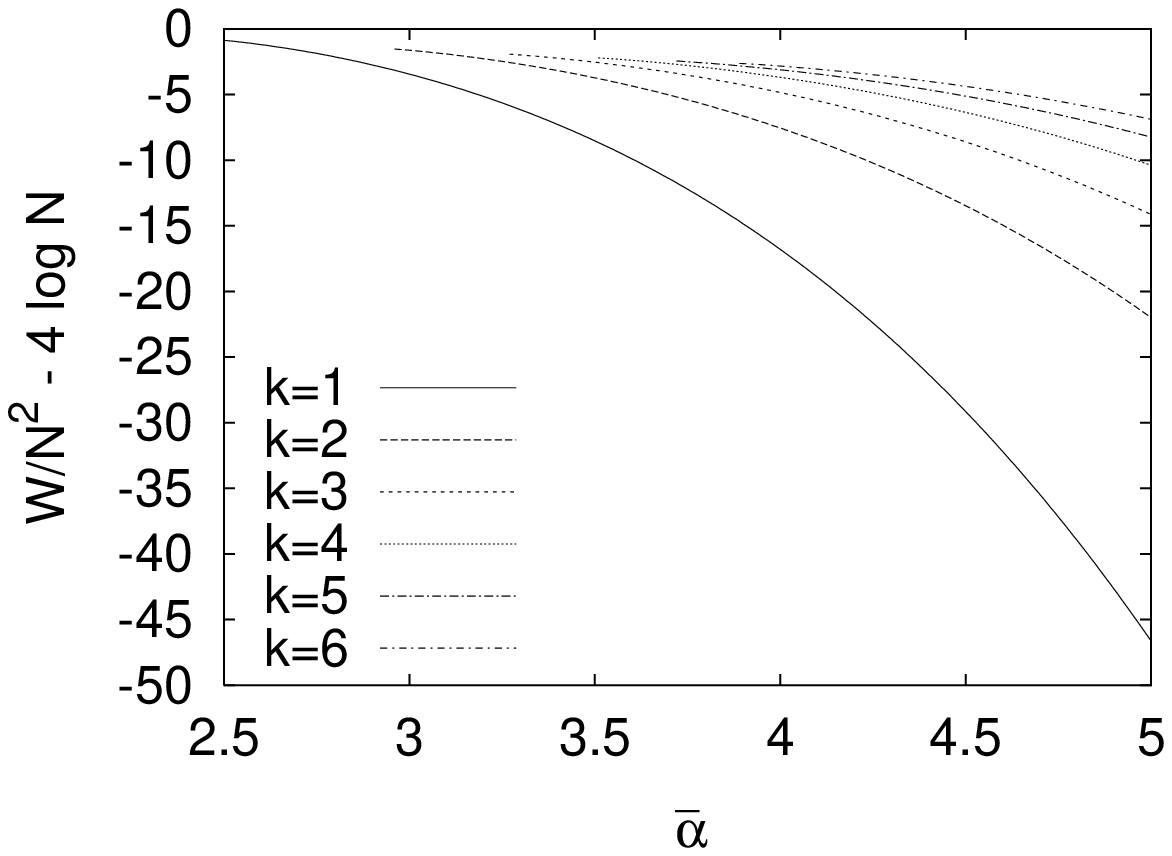,width=7.4cm}
    \epsfig{file=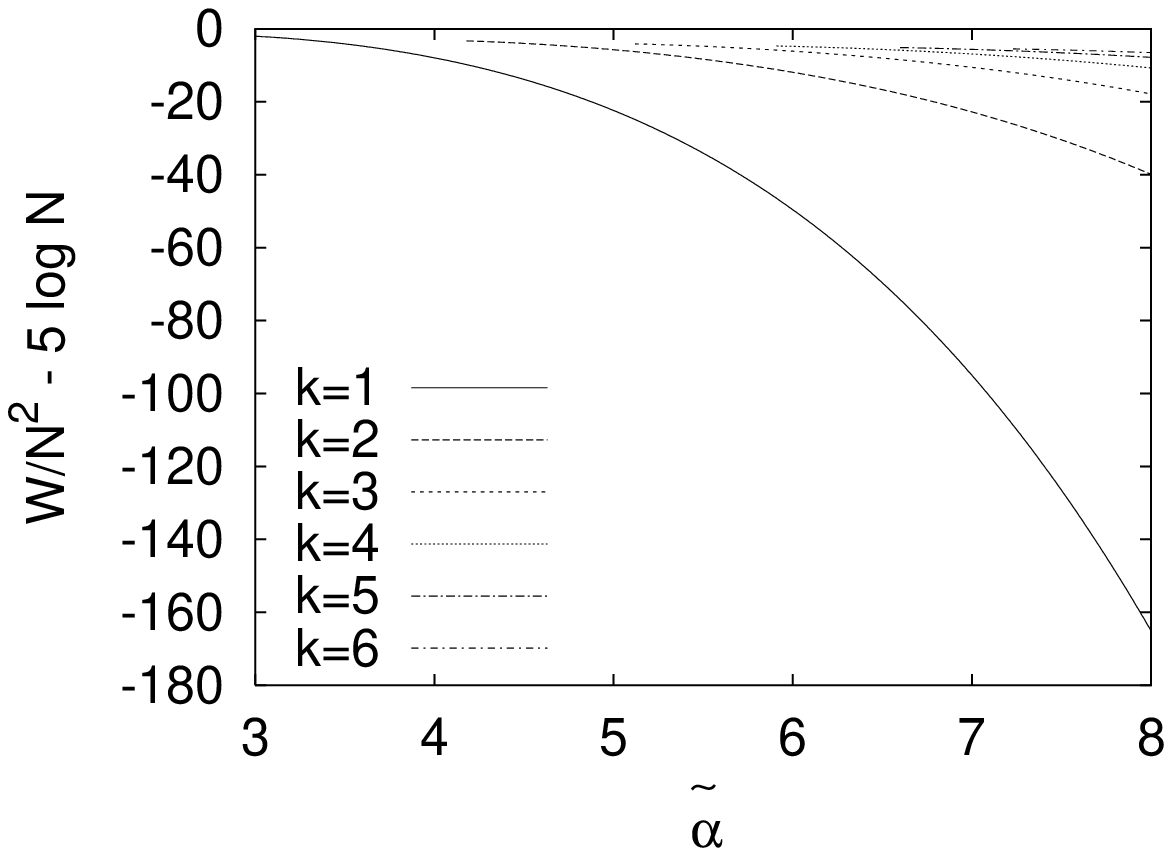,width=7.4cm}
   \caption{The free energy for the $k$ coincident fuzzy $\stwostwo$
   (left) and the $k$ coincident fuzzy $\stwo$ (right) 
   is plotted for $k=1,2,\cdots,6$.
   We have subtracted the irrelevant constants
   ($4 \log N$ and $5 \log N$, respectively) to make the quantity
   finite in the large-$N$ limit.} 
 \label{weff-k} }
 
 We therefore take $k=1$ and
 compare the free energy for the fuzzy $\stwostwo$ and the fuzzy $\stwo$.
 In order for the single fuzzy $\stwostwo$ to be stable, 
 we need to have
 ${\bar \alpha} > {\bar \alpha}^{(k=1 \, \stwostwo)}_{{\rm cr}}$.
 In that regime, however,  $\tilde \alpha \simeq \mbox{O}(N^{1/4})$ 
 and therefore
 $W_{k=1 \, \stwo} \simeq - \mbox{O}(N^3)$, 
 meaning that the single fuzzy $\stwo$
 has much smaller free energy than the single fuzzy $\stwostwo$.

 As we have done in ref.\ \cite{0504217},
 we can also calculate the free energy $W_{\rm YM}$ 
 in the Yang-Mills phase 
 for $\alpha \rightarrow 0$ in the large-$N$ limit
 since it should agree with the free energy 
 for the pure Yang-Mills model ($\alpha = 0$)
 studied in ref.\ \cite{Nishimura:2002va}.
 We obtain (See Appendix \ref{free-YM}.) 

\beq
\frac{1}{N^2} W_{\rm YM} = 3 \log N + 
 6 \left\{ -0.33 +  \log 
\left( 3^\frac{1}{4} \pi ^{-\frac{1}{2}} \right)
\right\} \ .
\eeq
 Comparing this result with 
 (\ref{s2s2-s2eff1loop}) for $k=1$,
 we find that the true vacuum of this model is given by the 
 single fuzzy $\stwo$ for $\tilde{\alpha} > \tilde{\alpha}_{\rm cr}$
 and by the pure 
 Yang-Mills vacuum for $\tilde{\alpha} < \tilde{\alpha}_{\rm cr}$,
 where the critical point is
\beq
\tilde{\alpha}_{\rm cr} \simeq \Bigl( 48 \log N \Bigr)^{1/4} \ .
\eeq
This result is {\em exact} at large $N$ since the calculation
of $W_{k=1 \, \stwo}$ and $W_{\rm YM}$ are both reliable
at the critical point $\tilde{\alpha}_{\rm cr}$.

 Note that the {\em lower} critical point 
 $\tilde{\alpha}_{\rm cr}^{\rm (l)}
  \equiv \tilde{\alpha}_{\rm cr}^{(k=1 \, \stwo)}$,
 at which the single fuzzy $\stwo$
 becomes unstable, is below $\tilde{\alpha}_{\rm cr}$.
 On the other hand, from Monte Carlo simulations
 we find that the {\em upper} critical point, at which the pure 
 Yang-Mills 
 vacuum becomes unstable, is $\alpha_{\rm cr} ^{\rm (u)}\simeq 1.51$,
 which is above $\alpha_{\rm cr}\equiv
 \frac{1}{\sqrt{N}}\tilde{\alpha}_{\rm cr}$.
 The existence of such three kinds of critical points is typical to
 first-order phase transitions. In the region 
 $ \alpha_{\rm cr}^{\rm (l)} < \alpha < \alpha_{\rm cr}^{\rm (u)}$
 we observe a hysteresis behavior in simulations
 similarly to the one observed in ref.\ \cite{0401038}.

  \FIGURE{
\epsfig{file=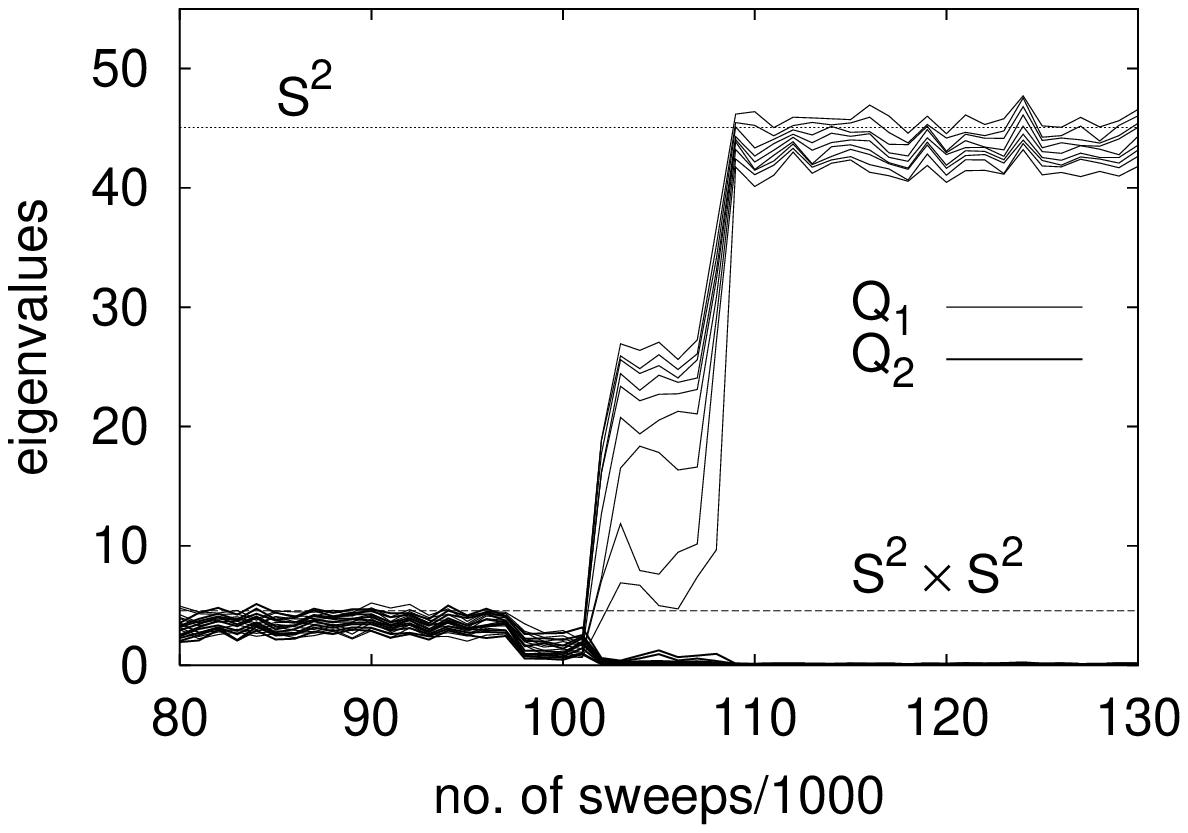,width=7.4cm}
 \caption{The eigenvalues of the ``Casimir operators'' 
$Q_{1}$ and $Q_{2}$ are plotted
  against the number of the sweeps in Monte Carlo simulation
 with the fuzzy $\stwostwo$ start.
 The dashed (dotted) line represents the classical value for the
fuzzy $\stwostwo$ (the fuzzy $\stwo$).
}
 \label{s2s2-decay} }

\section{A transition from the fuzzy $\stwostwo$ to the fuzzy $\stwo$}
\label{decay}

For $\alpha > \alpha_{\rm cr}$,
since the true vacuum is given by the fuzzy $\stwo$, 
we may be able to see 
the fuzzy $\stwostwo$ prepared as the initial configuration
decay into the fuzzy $\stwo$
in Monte Carlo simulations if we make a very long run.

In order to distinguish the classical solutions, it is convenient
to look at the eigenvalues of 
the ``Casimir operators'' $Q_{1}$ and $Q_{2}$.
In figure \ref{s2s2-decay}
we plot the eigenvalues obtained for $N=9$ and $\bar \alpha = 2.6$
(barely above ${\bar \alpha}^{(k=1 \, \stwostwo)}_{{\rm cr}}$)
against the number of the sweeps of the heat-bath algorithm
taking $A_{\mu} = X_\mu^{(n,n;1)}$ as the initial configuration.
For the first 97,000 sweeps all the eigenvalues of 
$Q_{1}$ and $Q_{2}$ are fluctuating around
the classical value $(R_1)^2 = (R_2)^2 \simeq 4.5$
for the symmetric fuzzy $\stwostwo$, but after a while
the eigenvalues of $Q_{1}$ and $Q_{2}$ 
are split into the classical values $(R_1)^2\simeq 45$ and 
$(R_2)^2 \simeq 0$ for the fuzzy $\stwo$,
which indicates that the configuration has become
 the single ($k=1$) fuzzy $\stwo$.
 In this particular event we observe that the fuzzy $\stwo$ is formed in the
 123-directions, but the formation of the fuzzy $\stwo$ in 
 the 456-directions should be observed with equal probability due to the 
 Z$_2$ symmetry for exchanging the two sets of directions.

 We have performed a simulation for $N=16$ with the same
 $\bar \alpha = 2.6$ starting again from the fuzzy $\stwostwo$,
 but the decay is not observed within 10,000,000 sweeps.
 This suggests that the potential barrier between the 
 fuzzy $\stwostwo$ and the fuzzy $\stwo$ grows with $N$, and therefore
 the decay probability is suppressed at large $N$.


\acknowledgments
We would like to thank Yoshihisa Kitazawa for helpful discussions. 
The work of T.A. and J.N.\ 
is supported in part by Grant-in-Aid for 
Scientific Research (Nos.\ 03740 and 14740163, respectively)
from the Ministry of Education, Culture, Sports, Science and Technology. 

\bigskip 

\appendix

\section{Evaluation of the free energy around a classical solution}
\label{one-loop-freeE}

In this section we formulate the perturbation theory
and derive a formula for the free energy.
Let us evaluate the partition function
$Z = \int d A \, e^{-S}$ around the
fuzzy $\stwostwo$ solution $X_\mu^{(n_1,n_2;k)}$
at the one-loop level.
The measure for the path integral is defined by
$dA = \prod_{\mu =1}^6 \prod_{a=1}^{N^2-1} \, dA_\mu^a $,
where $A_\mu = \sum_{a=1}^{N^2-1} A_\mu ^a \, t^a$
with $t^a$ being the generators of SU($N$) normalized by
$\tr (t^a \, t^b)=\delta_{ab}$.


We need to fix the gauge since there are 
flat directions corresponding to the transformation
$A_\mu \mapsto A_\mu^{g} \equiv 
g \, A_\mu \,  g^\dag$, where $g$ is an element
of the coset space 
$H \equiv {\rm U}(N)/ {\rm U}(k)$.
In order to remove the associated zero modes, 
we introduce the gauge fixing term and the
corresponding ghost term
\begin{eqnarray}
S_{\rm g.f.} &=& -\frac{N}{2}\text{tr}[X_\mu, A_\mu]^2 \ , \\
S_{\rm ghost}&=& -N\text{tr} \left([X_\mu,\bar{c}][A_\mu,c] \right) \ ,
\end{eqnarray}
where $c$ and $\bar c$ are the ghost and anti-ghost fields, respectively.

We perform the integration over $A_{\mu}$ perturbatively
by decomposing it as
$ A_{\mu} = X_{\mu} + {\tilde A}_{\mu}$,
where $X_\mu \equiv X_\mu^{(n_1,n_2;k)}$.
The partition function can be written as
 \begin{eqnarray}
Z =  \mbox{vol}(H) \, {\cal N}
\int  \dd \tilde A \, \dd c 
\, \dd \bar c \, \ee ^{-S_{\rm total}}  \ ,
\label{part-perturbation}
 \end{eqnarray}
where the total action 
$S_{\rm total} = S + S_{\rm g.f.} + S_{\rm ghost}$
is given
by
\beqa
S_{\rm total} &=& S[X]+ S_{\rm kin} + S_{\rm int}  \ ,\\
S_{\rm kin}&=&\frac{1}{2}\,  N \, \tr \left( 
\tilde A_\mu [X_\lambda , [X_\lambda , \tilde A_\mu ]] \right)
+ N \, \tr \Bigl( \bar c \, [X_\lambda , [X_\lambda , c ]] \Bigr) \ ,
%
%
\label{SUalg_term}
\\
S_{\rm int}&=&-N \, \tr \left( 
[\tilde A_{\mu} , \tilde A_{\nu} ][X_{\mu} ,\tilde A_{\nu} ] \right) 
- \frac{1}{4}\,  N\, 
\tr \left( [\tilde A_{\mu} , \tilde A_{\nu} ]^2 \right) \non
&&+ \frac{2}{3}\, i \, \alpha \, f_{\mu \nu \rho} \, N \,
 \tr \left( \tilde A_{\mu}  \tilde A_{\nu}  \tilde A_{\rho}  \right)
+ N \, \tr \left( \bar c \, [X_{\mu}, [\tilde A_{\mu} , c]] \right) \ .
\eeqa
The volume of the coset space $H$, 
$\mbox{vol}(H)= \mbox{vol}({\rm U}(N))/ 
\mbox{vol}({\rm U}(k))$,
can be obtained by the formula
$\mbox{vol}({\rm U}(p)) = 
\frac{(2 \pi)^{\frac{p(p+1)}{2}}}{(p-1)\, ! \cdots 0 \, !  }$,
and the normalization factor 
${\cal N} = (2 \pi N)^{- \frac{1}{2}(N^2 - k^2)}$
can be obtained by following the usual gauge fixing procedure as in
ref.\ \cite{0504217}.
The prefactors $\mbox{vol}(H)$ and ${\cal N}$ in
(\ref{part-perturbation}) are omitted in
refs.\ \cite{0401038,0410263} since they are irrelevant for
the discussions in those papers, 
but we need to keep them for the comparison with 
the free energy in the Yang-Mills phase discussed in
section \ref{true}.

We calculate the free energy
$W = - \log Z$ as a perturbative expansion 
$W = \sum_{j=0}^{\infty} W_j$,
where $W_j = \mbox{O}(\alpha^{4(1-j)})$.
The classical part $W_{0}$ is given by eq.\ (\ref{cl-action}).
Note that the kinetic term (\ref{SUalg_term}) can be written as
\beq
S_{\rm kin} =
N  \tr\left\{
\frac{1}{2}  \, \tilde A_\mu  ({\cal P}_\lambda )^2  \tilde A_\mu  
 +  \bar c \, ({\cal P}_\lambda )^2  c \right\} \ ,
\label{SQ_fs}  
\eeq
where we have introduced the operator ${\cal P}_{\mu}$
\begin{equation}
{\cal P}_{\mu} M \defeq [X_{\mu}, M]  \ ,
\end{equation}
which acts on the space of $N \times N$ traceless matrices.
The one-loop term is obtained as
\beq
W_1 = 2 \, {\cal T}r ' \log \left\{ N ({\cal P}_\mu)^2 \right \}  
- \log \Bigl\{  \mbox{vol}(H) \, {\cal N} \Bigr\}
\ , \label{one-loop_Seff}
\eeq
where the symbol ${\cal T}r ' $ denotes
the trace in the space of $N \times N$ matrices
omitting the zero modes\footnote{
Strictly speaking, we have to treat the zero modes for $k\neq 1$
more carefully. See ref.\ \cite{0401038} for a discussion
on this point. This complication, however, 
does not affect our conclusion 
concerning the large-$N$ limit with fixed $k$.}.
 

\section{Perturbative calculations for the fuzzy $\stwostwo$}
\label{one-loop-perturbation}

In this section we consider the symmetric fuzzy $\stwostwo$
taking $n_1=n_2 =n (= \sqrt{\frac{N}{k}})$ for the solution $X_\mu=X_\mu^{(n_1,n_2;k)}$.

\subsection{One-loop calculation of the free energy}
\label{eff_derive}
In order to solve the eigenvalue problem 
of the operator $({\cal P}_\lambda)^2$, let us introduce
\beq
Y_{lm}^{(a,b)} \defeq
\Bigl(\hat{Y}_{l_1,m_1} \otimes 
\hat{Y}_{l_2,m_2} \Bigr) \otimes {\bf e}^{(a,b)} \ ,
\eeq
where $\hat{Y}_{lm}$ represents
$n$-dimensional matrix spherical harmonics,
and ${\bf e}^{(a,b)}$ is a $k\times k$ matrix 
whose ($a,b$) element is 1 and all the other elements are zero.
The indices $l$ and $m$ represent the double indices
$(l_1 , l_2)$ and $(m_1 , m_2)$, respectively.
The matrices $Y_{lm}^{(a,b)}$
form a complete basis of $N \times N$ matrices, and they
have the properties
\beqa
\tr \left( Y_{lm}^{(a,b)\dag} \,  Y_{l'm'}^{(a',b')}  \right)
&=& \delta_{ll'}\delta_{mm'}\delta_{aa'}\delta_{bb'} \ ,\\
Y_{lm}^{(a,b)\dag} &=& (-1)^{m_1}(-1)^{m_2} Y_{l,-m}^{(b,a)} \ .
\eeqa
Acting the operator $({\cal P}_\lambda)^2$ on $Y_{lm}^{(a,b)}$, we find
\begin{equation}
({\cal P}_\mu)^2 Y_{lm}^{(a,b)}=
\alpha^{2} \left\{ l_1(l_1+1) + l_2(l_2+1) \right\} Y_{lm}^{(a,b)} \ ,
\end{equation}
which means that $Y_{lm}^{(a,b)}$ are the eigenstates of 
$({\cal P}_\lambda)^2$.
Thus the first term in eq.\ (\ref{one-loop_Seff}) can be obtained as
\begin{eqnarray}
2 \, {\cal T}r ' \log \left\{ N ({\cal P}_\mu)^2 \right \} 
&=&  2 k^{2} \sum _{l_1, l_2=0}^{n-1} 
\hspace{-2mm} 
{\bf '} 
\hspace{2mm}
(2l_1+1)(2l_2+1) 
\log \left[ N \alpha^{2} \left\{l_1(l_1+1) + l_2(l_2+1) \right\} \right] 
 \n  \\
&\simeq&
N^{2} \left(
\log \frac{16N^3}{k^2} - 3 + 4 \log {\bar \alpha} \right) \ ,\label{1st-term}
\end{eqnarray}
where the large-$N$ limit is taken in the second line with fixed $k$,
and the symbol $\sum \, '$ implies that the zero modes
$l_1=l_2=0$ are excluded.
The second term in eq.\ (\ref{one-loop_Seff}) can be obtained as
\begin{eqnarray}
- \log \Bigl\{  \mbox{vol}(H) \, {\cal N} \Bigr\}
\simeq N^{2} \left(
\log N - \frac{3}{4}  \right) \ .\label{2nd-term}
\end{eqnarray}
Adding the two terms, 
we obtain the one-loop contribution $W_{1}$ as
\begin{eqnarray}
W_1 &=& 
N^{2} \left(
\log \frac{16N^4}{k^2} - \frac{15}{4} 
+ 4 \log {\bar \alpha} \right) \ .\label{coin_ol}
\end{eqnarray}


\subsection{One-loop calculation of various observables}


In order to calculate various observables in perturbation
theory, we need the propagators for $\tilde A_\mu$ and the ghosts,
which are given as
\beqa
\left\langle (\tilde A_\mu)_{ij} (\tilde A_\nu)_{kl}
\right\rangle _0
&=& 
\delta_{\mu\nu} \sum_{ab}
\sum _{l_1, l_2=0}^{n-1} 
\hspace{-2mm} 
{\bf '} 
\hspace{2mm}
 \sum_{m_1=-l_1}^{l_1}
 \sum_{m_2=-l_2}^{l_2} \n \\
&&
\frac{(-1)^{m_1} (-1)^{m_2}}
{N \alpha^{2} \left\{ l_1(l_1 +1) + l_2(l_2 +1)\right\}}
\left( Y_{lm}^{(a,b)}  \right)_{ij}
\left( Y_{l, -m}^{(b,a)} \right)_{kl} \ , \\
\Bigl \langle(c)_{ij} (\bar c)_{kl} \Bigr\rangle_0 &=& 
\sum_{ab}
\sum _{l_1, l_2=0}^{n-1} 
\hspace{-2mm} 
{\bf '} 
\hspace{2mm}
\sum_{m_1=-l_1}^{l_1}
\sum_{m_2=-l_2}^{l_2} \n \\
&&
\frac{(-1)^{m_1} (-1)^{m_2}}
{N \alpha^{2} \left\{ l_1(l_1 +1) + l_2(l_2 +1)\right\}}
\left( Y_{lm}^{(a,b)}  \right)_{ij}
\left( Y_{l, -m}^{(b,a)} \right)_{kl} \ , 
\eeqa
where the symbol $\langle \ \cdot \ \rangle_0$ refers to the
expectation value calculated using the kinetic term $S_{\rm kin}$ 
in (\ref{SUalg_term}) only. 

We also need to obtain the tadpole
$\langle \tilde A_\mu \rangle$, which can be expressed as 
\beq
\langle\tilde A_\mu \rangle = c \, X_{\mu} 
\eeq
with some coefficient $c$ due to the SO($3$)$\times$SO($3$)$\times$
Z$_2$ symmetry. Since
\begin{eqnarray}
\frac{1}{N} \tr \sum_{\mu=1}^{3} ( X_\mu \langle \tilde A_\mu \rangle )
 &=& c \, \tr \sum_{\mu=1}^{3} (X_\mu X_\mu)
= \frac{c}{4} \, \alpha^2 \, (n^2 -1 )  \ ,
\label{inner-product1}
\end{eqnarray}
we can obtain the coefficient $c$ by evaluating the left most term
in (\ref{inner-product1}). 
At the leading order in $\frac{1}{\alpha^4}$, we obtain
\begin{eqnarray}
\frac{1}{N} \tr \sum_{\mu=1}^{3}
\left( X_\mu \langle \tilde A_\mu \rangle_{\rm 1-loop} \right)
&=& 
 \left\langle \tr \sum_{\mu=1}^{3} \sum_{\nu,\rho=1}^{6} 
(X_\mu \tilde A_\mu ) \, 
\tr \left([\tilde A_{\nu} ,\tilde A_{\rho} ][X_{\nu} ,\tilde
A_{\rho} ] \right) \right\rangle_0 \n \\
&~& 
- \left\langle \tr \sum_{\mu=1}^{3} \sum_{\nu,\rho,\sigma=1}^{6}
 (X_\mu \tilde A_\mu)
 \,  \tr\left( \frac{2}{3}\, i \, \alpha \, 
f_{\nu \rho \sigma} \tilde A_{\nu}  \tilde A_{\rho}  \tilde
A_{\sigma}  \right)\right\rangle_0 \n \\
&~&  -  \left\langle \tr \sum_{\mu=1}^{3} \sum_{\nu=1}^{6}
(X_\mu \tilde A_\mu)  \, \tr
\left( \bar c \, [X_{\nu} ,[\tilde A_{\nu} ,c] ] \right) 
\right\rangle_0 \ .
\label{tadpole}
\end{eqnarray}
Using the fact that $L_\mu^{(n)}$ can be 
written as a linear combination of $\hat{Y}_{l=1,m}$, 
we can evaluate (\ref{tadpole}) as
\begin{eqnarray}
 \tr \sum_{\mu=1}^{3} 
 \left(X_\mu\langle\tilde A_\mu \rangle_{\rm 1-loop}\right)
&=&
-\frac{2}{N \alpha^{2}} k^2 
\sum _{l_1, l_2=0}^{n-1} 
\hspace{-2mm} 
{\bf '} 
\hspace{2mm}
\frac{(2l_1 +1)(2l_2 +1) l_1 (l_1+1)}
{l_1 (l_1+1) + l_2 (l_2+1)} \  .
\label{tr_L^red_A1}
%
\end{eqnarray}
{}From (\ref{inner-product1}) we obtain
\begin{eqnarray}
c &=& - \frac{8k^2}{N^2 \alpha^4 (n^2-1)}
\sum _{l_1, l_2=0}^{n-1} 
\hspace{-2mm} 
{\bf '} 
\hspace{2mm}
\frac{(2l_1 +1)(2l_2 +1) l_1 (l_1+1)}
{l_1 (l_1+1) + l_2 (l_2+1)}  \ .
%
\end{eqnarray}
%


Using the propagator and the tadpole obtained above,
we can evaluate various observables at the one-loop level.
For instance, the ``extent of space-time'' is given by
\beqa
\left\langle \frac{1}{N} \tr (A_\mu)^2 \right\rangle_{\rm 1-loop} 
&=&
\frac{1+2c}{2} \alpha^2 (n^2 -1) \nonumber \\
&~& +
\sum _{l_1, l_2=0}^{n-1} 
\hspace{-2mm} 
{\bf '} 
\hspace{2mm}
\frac{ 6 k^2 (2l_1 +1) (2l_2 +1)}
{N^2 \alpha^{2} \left\{l_1(l_1+1) + l_2(l_2 +1)\right\}} \ .
\label{trA2-form}
\eeqa
At large $N$ with fixed 
$\bar \alpha \equiv \alpha N^{\frac{1}{4}}$, we obtain
\begin{eqnarray}
\frac{1}{\sqrt{N}}
\left\langle \frac{1}{N} \tr (A_\mu)^2 \right\rangle_{\rm 1-loop} 
\simeq 
\frac{\bar\alpha^2}{2k}
- \frac{4}{\bar\alpha^2} \ . \label{tr_A^2_largeN}
\end{eqnarray}

The expectation value $\langle S \rangle$ can 
be calculated in a similar manner,
but it is much easier to calculate it in the following way.
Let us define a rescaled action
\begin{eqnarray}
 S(\lambda, \alpha) = \lambda \, 
N \, \tr \left( - \frac{1}{4} \,[A_{\mu}, A_{\nu}]^{2}
  + \frac{2}{3}  \, i \,  \alpha \, f_{\mu \nu \rho} \,
 A_{\mu} A_{\nu} A_{\rho} \right) 
\end{eqnarray}
and the corresponding free energy 
\begin{eqnarray}
 e^{-W(\lambda, \alpha)} = \int d A \, e^{-S(\lambda, \alpha)} \ ,
\end{eqnarray}
which is related to the original free energy $W = W(1, \alpha)$ through
\begin{eqnarray}
 W(\lambda, \alpha) = \frac{3}{2} \, (N^{2}-1) \, \log \lambda 
 + W(1, \lambda^{\frac{1}{4}} \alpha) \ .
\end{eqnarray}
Then we obtain the expectation value $\langle S \rangle$ as
\begin{eqnarray}
 \frac{\langle S \rangle}{N^{2}} = \left.
\frac{1}{N^{2}} \frac{\partial W(\lambda, \alpha)}{\partial
 \lambda} \right|_{\lambda=1} &=& 
\frac{3}{2} \, \left(1 - \frac{1}{N^{2}}\right) 
+ \frac{1}{4N^{2}} \, {\bar \alpha} \, 
 \frac{\partial W}{\partial {\bar \alpha}} 
\label{one-loopacto2}
\\ &\simeq&  
 - \frac{{\bar \alpha}^{4}}{12k} + \frac{5}{2} \ .
\end{eqnarray}

\subsection{Critical point}

As we see in section \ref{stwostwo},
Monte Carlo simulations show that 
the fuzzy $\stwostwo$ decays into the pure Yang-Mills vacuum
at some critical point.
We can reproduce the critical point by 
perturbative calculations as follows.
Let us consider the one-loop effective action
for the rescaled fuzzy $\stwostwo$ configuration
\begin{eqnarray}
A_{\mu} = 
\left\{
\begin{array}{l}
\beta \, (L_\mu^{(n)}\otimes {\bf 1}_{n}) 
\otimes {\bf 1}_{k} 
\quad {\rm for} \quad \mu =1,2,3  \ , \\
\beta \,  ( {\bf 1}_{n} \otimes L_\mu^{(n)} ) 
\otimes {\bf 1}_{k} 
\quad {\rm for} \quad \mu =4,5,6 \ ,
\end{array}
\right.
\label{fsb-k}
\end{eqnarray}
which is given at large $N$ as
  \begin{eqnarray}
\frac{1}{N^2} \, 
   \Gamma_{k \, \stwostwo} (\bar \beta) &\simeq& 
   \left( \frac{1}{8}\, {\bar \beta}^{4} - 
\frac{1}{6} \, {\bar \alpha} {\bar \beta}^{3} \right) 
    \frac{2}{k} + 4 \,  \log {\bar \beta} 
 + \log \frac{16N^4}{k^2} - \frac{15}{4} 
  \ , \label{eff-s2s2-1loop}
  \end{eqnarray}
  where ${\bar \beta}  = \beta  N^{\frac{1}{4}}$.
The derivation is similar to the calculation
of the free energy described in
section \ref{eff_derive}. 
(Note that one-particle reducible diagrams
do not appear at the one-loop level.)
In fact the effective action is
one-loop exact in the large-$N$ limit as can be shown
by a power counting argument \cite{0307007}.
The effective action has extrema at $\bar \beta$ satisfying
  \begin{eqnarray}
\frac{1}{N^2} \, 
\frac{\partial \Gamma_{k \, \stwostwo} (\bar\beta)}
{\partial {\bar \beta}}
   =  \frac{1}{k} 
    ({\bar \beta}^{3} - {\bar \alpha} {\bar \beta}^{2})
   +  \frac{4}{{\bar \beta}} = 0 \ . \label{criticalpt}
  \end{eqnarray}
{}From this we find that the
effective action $\Gamma_{k \, \stwostwo}$ has
a local minimum if and only if 
$\bar \alpha >
 {\bar \alpha}^{(k \, \stwostwo)}_{{\rm cr}} $, where
the critical point $ {\bar \alpha}^{(k \, \stwostwo)}_{{\rm cr}} $
is given by
  \begin{eqnarray}
 {\bar \alpha}^{(k \, \stwostwo)}_{{\rm cr}} 
 = \frac{4}{3} \times (12k)^{\frac{1}{4}}
   \simeq 2.481613 \cdots \times k^{\frac{1}{4}} \ .  
   \label{kuri}
  \end{eqnarray}

\subsection{All order results}
 \label{allorder-s2s2}

 Since the free energy and the effective action
 are related to each other by the Legendre transformation, we can
 obtain the free energy by evaluating the effective action at its 
 extremum. Then exploiting the fact that 
 the effective action is one-loop exact,
 we can evaluate the free energy to all orders
 in perturbation theory.
 This method is proposed in ref.\ \cite{0403242}, and it is applied
 to other models successfully \cite{0405277,0410263}. 

 Above the critical point 
 ${\bar \alpha} > {\bar \alpha}^{(k \, \stwostwo)}_{{\rm cr}}$, 
  the value of ${\bar \beta}$ that gives the local minimum
  of the effective action
 can be obtained by solving eq.\ (\ref{criticalpt}) as
  \begin{eqnarray}
   {\bar \beta} &=& f(\bar \alpha) \defeq
  \frac{\bar \alpha}{4} \left( 1 + \sqrt{1+ \delta} + \sqrt{2 - \delta
   + \frac{2}{\sqrt{1 + \delta}} }  \right) \ , 
  \label{beta-crit} 
  \end{eqnarray}
  where
  \begin{eqnarray}
   \delta = 
   {\bar \alpha}^{-\frac{4}{3}} (128k)^{\frac{1}{3}} \left\{ \left( 1 
   + \sqrt{1 - \frac{1024 k}{27 {\bar \alpha}^{4}}} \right)^{\frac{1}{3}}
   + \left(1 - \sqrt{1 - \frac{1024 k}{27 {\bar \alpha}^{4}}} 
   \right)^{\frac{1}{3}}
   \right\} \ .
  \end{eqnarray}
At large ${\bar \alpha}$, the solution (\ref{beta-crit}) is expanded as
 \begin{eqnarray}
   {\bar \beta} =  f(\bar \alpha) 
  = {\bar \alpha} \left( 1 - \frac{4k}{{\bar \alpha}^{4}} - \frac{48k^{2}}
  {{\bar \alpha}^{8}}
  - \frac{960 k^{3}}{{\bar \alpha}^{12}} - \cdots \right) \ .
  \label{beta-crit2}
 \end{eqnarray}
Plugging this solution into (\ref{eff-s2s2-1loop}), 
we obtain the free energy to all orders as
 \begin{eqnarray}
\frac{1}{N^2} \,   W_{k \, \stwostwo} 
\simeq - \frac{{\bar \alpha}^{4}}{12k} 
+ 4 \log {\bar \alpha}
+ \log \frac{16N^4}{k^2} - \frac{15}{4} 
 - \frac{8k}{{\bar \alpha}^{4}} 
  - \frac{224k^{2}}{3{\bar \alpha}^{8}}
  - \frac{3520 k^{3}}{3 {\bar \alpha}^{12}} 
  - \cdots \ . \mbox{~~~} 
 \label{w-exact}
 \end{eqnarray}
 Using (\ref{one-loopacto2}), we obtain the all order result for
 the expectation value $\langle S \rangle$ as
   \begin{eqnarray}
  \frac{1}{N^{2}} \, \langle S \rangle
\simeq - \frac{{\bar \alpha}^{4}}{12k} + \frac{5}{2}
  + \frac{8k}{{\bar \alpha}^{4}} + \frac{448 k^{2}}{3{\bar \alpha}^{8}}
  + \frac{3520 k^{3}}{{\bar \alpha}^{12}} + \cdots \ .
    \label{acto-exact}
 \end{eqnarray}

 We can also calculate other observables
 to all orders as in ref.\ \cite{0410263}.
 For instance, let us calculate 
 $\left\langle \frac{1}{N} \tr (A_\mu)^2 \right\rangle$.
 Since the one-loop contribution comes totally
 from the tadpole diagrams, which are one-particle reducible
 (The second term in eq.\ (\ref{trA2-form}) is subleading in $N$.),
 we may obtain the all order result by simply
 replacing ${\bar \alpha}$ by the quantum solution $f(\bar \alpha)$
 in the classical result as
    \begin{eqnarray}
\frac{1}{\sqrt{N}} \left\langle \frac{1}{N} \tr (A_\mu)^2 \right\rangle
  &\simeq& \frac{1}{2k} \, f({\bar \alpha})^{2}
   = \frac{{\bar \alpha}^{2}}{2k} - \frac{4}{{\bar \alpha}^{2}}
   - \frac{40k}{{\bar \alpha}^{6}}
   - \frac{768k^{2}}{{\bar \alpha}^{10}} 
   - \frac{18304k^{3}}{{\bar \alpha}^{14}} - \cdots \ . \mbox{~~~}
    \label{a-sq-exact}
 \end{eqnarray}

\section{Perturbative calculations for the fuzzy S$^{2}$}
\label{s2oneloop}
In this section we briefly describe the perturbative calculations
for the fuzzy ${\rm S}^2$ 
taking $n_1=n (= \frac{N}{k})$ and $n_2=1$ for the solution $X_\mu=X_\mu^{(n_1,n_2;k)}$.
%

\subsection{One-loop calculations}

The one-loop free energy is given at large $N$ by
\begin{eqnarray}
\frac{1}{N^2} \,    W_{k \, \stwo} 
&=& -\frac{\alpha^{4} N^2}{24}(n^2-1)
+ 2 k^2 \sum_{l=1}^{n-1} 
(2l+1)\log[N \alpha^{2} l(l+1)] 
- \log \Bigl\{  \mbox{vol}(H) \, {\cal N} \Bigr\}
\n \\
&\simeq& N^2 \left( -\frac{{\tilde \alpha}^4}{24k^2}
+ 4 \log{\tilde \alpha}
+  \log{\frac{N^5}{k^4}} - \frac{11}{4}
\right) \ ,
\label{OL-effact_S2}
\end{eqnarray}
where $\tilde\alpha = \alpha N^{\frac{1}{2}}$. 
The tadpole is given by
\begin{eqnarray}
\langle \tilde A_\mu \rangle_{\rm 1-loop}= 
\left\{ \begin{array}{ll} 
 -\frac{8k^2}{N^2 \alpha^4}X_\mu  
 & \textrm{ for $\mu=1,2,3$}  \ , \\ 
0 & \textrm{ for $\mu=4,5,6$} \ .
\end{array} \right. 
\end{eqnarray}
The observables
can be evaluated as
\begin{eqnarray}
\frac{1}{N}
\left\langle \frac{1}{N} \, \tr \, Q_1 \right\rangle
_{\rm 1-loop} 
&=&
\frac{1}{N^2}\Bigl\{ \tr(X_\mu X_\mu)
+  2 \, \tr \left( X_\mu \langle 
\tilde A_\mu\rangle _{\rm 1-loop} \right)
+\langle\tr (\tilde A_\mu)^2 \rangle_0 \Bigr\} \n \\
&=&
\alpha^2 \left\{
\frac{1}{4N}  (n^2 -1) 
-\frac{4k^2(n^2-1)}{N^3 \alpha^4}  
+\frac{6}{N n^2 \alpha^4} \sum_{l=1}^{n-1}\frac{2l+1}{l(l+1)}
\right\} \n \\
&\simeq&
\frac{{\tilde \alpha}^{2}}{4k^2}-\frac{4}{\tilde\alpha^2}  \ , \\
\frac{1}{N}
\left\langle \frac{1}{N} \, \tr \, Q_2 \right\rangle
_{\rm 1-loop} 
&\simeq& 0 \ , \\
\frac{1}{N^{2}} \langle S \rangle_{\rm 1-loop} 
&=&  -\frac{\alpha^4}{24} (n^2 -1) + \frac{5}{2}
+\frac{1}{4N^2}\Bigl\{ 6(-k^2-1) + 2k^2 \Bigr\} \n \\
&\simeq& 
-\frac{\tilde\alpha^4}{24k^2} + \frac{5}{2} \ .
\end{eqnarray}

\subsection{Critical point}

The critical point can be obtained by
considering the effective action for the rescaled fuzzy
$\stwo$ configuration
  \begin{eqnarray}
   A_{\mu} = \left\{ \begin{array}{ll} 
    \beta (L^{(n)}_{\mu} \otimes {\bf 1}_{k}) & \mbox{~~~for~}\mu=1,2,3 \ , \\
    0 & \mbox{~~~for~}\mu=4,5,6 \  . \end{array} \right.
  \label{fs2b-k} 
  \end{eqnarray}
The effective action is obtained in the large-$N$ limit as
\begin{eqnarray}
\frac{1}{N^{2}}
 \Gamma_{k \, \stwo} (\tilde\beta)
 &\simeq&   \left( 
\frac{1}{8}\, {\tilde \beta}^{4} 
- \frac{1}{6} \, {\tilde \alpha} {\tilde \beta}^{3} \right)
 \frac{1}{k^{2}}
 + 4 \, \log {\tilde \beta} 
 +  \log{\frac{N^5}{k^4}} - \frac{11}{4}  \ , 
\label{eff-s2-1loop} 
\end{eqnarray}
where ${\tilde \beta}  = \beta  \sqrt{N}$.
The critical point is obtained as
\begin{eqnarray}
   {\tilde \alpha}_{{\rm cr}}^{(k \, {\rm S}^{2})} = \frac{4}{3} \times 
   24^{\frac{1}{4}} \sqrt{k} 
   = 2.9511518 \cdots \times \sqrt{k} \ .
\label{anal-crit_S2}
\end{eqnarray}

\subsection{All order results}
\label{allorder-s2s2-s2}
Above the critical point 
$\tilde \alpha > {\tilde \alpha}_{{\rm cr}}^{(k \, {\rm S}^{2})}$,
the effective action has a local minimum at
 \begin{eqnarray}
   {\tilde \beta} &=& g(\tilde \alpha) \defeq
\frac{\tilde \alpha}{4} \left( 1 + \sqrt{1+ \varepsilon} + 
\sqrt{2 - \varepsilon
   + \frac{2}{\sqrt{1 + \varepsilon}} }  \right), \label{beta-crit-s2} \\
   \varepsilon  &=& 
{\tilde \alpha}^{-\frac{4}{3}} (256 k^{2})^{\frac{1}{3}} \left\{ \left( 1 
   + \sqrt{1 - \frac{2048 k^{2}}{27 {\tilde \alpha}^{4}}} \right)^{\frac{1}{3}}
   + \left(1 - \sqrt{1 - \frac{2048 k^{2}}
   {27 {\tilde \alpha}^{4}}} \right)^{\frac{1}{3}}
   \right\} \ .
  \end{eqnarray}
The all order result for the free energy can be obtained
by substituting $\tilde\beta = g(\tilde \alpha)$ 
in (\ref{eff-s2-1loop}).
\begin{eqnarray}
   \frac{1}{N^{2}}  W_{k \, \stwo} 
  &\simeq& 
- \frac{{\tilde \alpha}^{4}}{24k^{2}} + 4 \log {\tilde \alpha}
+  \log{\frac{N^5}{k^4}} - \frac{11}{4}
  - \frac{16k^{2}}{{\tilde \alpha}^{4}} - \frac{896k^{4}}{3{\tilde \alpha}^{8}}
  - \frac{28160 k^{6}}{3{\tilde \alpha}^{12}}
  - \cdots \ . \mbox{~~~}
  \label{w-exact-s2}
  \end{eqnarray}
 The expectation values of observables can be calculated at large $N$ as
  \begin{eqnarray}
    \frac{1}{N^{2}} \, \langle S \rangle
  &\simeq& - \frac{{\tilde \alpha}^{4}}{24k^{2}} + \frac{5}{2}
  + \frac{16k^{2}}{{\tilde \alpha}^{4}} + 
\frac{1792 k^{4}}{3{\tilde \alpha}^{8}}
  + \frac{28160 k^{6}}{{\tilde \alpha}^{12}} + \cdots, 
   \label{acto-exact-s2} \\
    \frac{1}{N} \left\langle \frac{1}{N} \, \tr \, 
     Q_{1} \right\rangle &\simeq&
   \frac{{\tilde \alpha}^{2}}{4k^{2}} - \frac{4}{{\tilde \alpha}^{2}}
  - \frac{80k^{2}}{{\tilde \alpha}^{6}} 
   - \frac{3072 k^{4}}{{\tilde \alpha}^{10}}
  - \frac{146432k^{6}}{{\bar \alpha}^{14}} - \cdots,
    \label{a-sq-exact-s2} \\
    \frac{1}{N} \left\langle \frac{1}{N} \, \tr \, Q_{2} \right\rangle 
    &\simeq& 0 \ .
 \end{eqnarray}

\section{Free energy in the Yang-Mills phase}
\label{free-YM}

In ref.\ \cite{Nishimura:2002va} the free energy for the pure Yang-Mills
model ($\alpha = 0$) is calculated as
\beq
\frac{1}{N^2} F \simeq
D
\left\{ f_D + \frac{1}{2} \ln \left( \sqrt{\frac{D}{2}}N \right) 
- \frac{1}{2} \ln \pi \right\} \ ,
\eeq
where $D$ is the number of bosonic matrices, which is $D=6$ in the
present case, and the third term is added to adjust the 
normalization of the measure to the one used in the present paper.
The ``free energy density'' $f_D$ can be calculated analytically by the
$1/D$ expansion \cite{9811220}, and one obtains
$f_{\infty} = -  1/4$ at $D=\infty$.
For $D=6$
the gaussian expansion method \cite{Nishimura:2002va}
gives $f_6=-0.33$ (the convergence is clear up to order 7).
Therefore the free energy in the Yang-Mills phase of our model is given
at large $N$ as
\beq
\frac{1}{N^2} W_{\rm YM} \simeq 3 \log N + 
6 \, \left\{ -0.33 
+  \log 
\left( 3^\frac{1}{4} \pi ^{-\frac{1}{2}} \right)
\right\} \ ,
\eeq
if $\alpha$ is sent to zero in the large-$N$ limit.
Finite $\alpha$ corrections should be O($\alpha^2$)
due to the parity symmetry of the pure Yang-Mills model.

\end{document}